\documentclass[%
 reprint,
 amsmath,amssymb,
 aps,preprintnumbers,
]{revtex4-1}

\usepackage{graphicx}
\usepackage{dcolumn}
\usepackage{bm}

\begin{document}

\preprint{FERMILAB-PUB-19-019-A}


\title{Constraining Sterile Neutrino Interpretations of the LSND and MiniBooNE Anomalies with Coherent Neutrino Scattering Experiments}

\author{Carlos Blanco$^{a,b}$}
\author{Dan Hooper$^{b,c,d}$}
\author{Pedro Machado$^{e}$}

\affiliation{$^a$University of Chicago, Department of Physics, Chicago, IL 60637}
\affiliation{$^b$University of Chicago, Kavli Institute for Cosmological Physics, Chicago, IL 60637}

\affiliation{$^c$Fermi National Accelerator Laboratory, Theoretical Astrophysics Group, Batavia, IL 60510}
\affiliation{$^d$University of Chicago, Department of Astronomy and Astrophysics, Chicago, IL 60637}

\affiliation{$^e$Fermi National Accelerator Laboratory, Theoretical Physics Department, Batavia, IL 60510}

\date{\today}

\begin{abstract}

Results from the LSND and MiniBooNE experiments have been interpreted as evidence for a sterile neutrino with a mass near the electronvolt scale. Here we propose to test such a scenario by measuring the coherent elastic scattering rate of neutrinos from a pulsed spallation source. Coherent scattering is universal across all active neutrino flavors, and thus can provide a measurement of the total Standard Model neutrino flux. By performing measurements over different baselines and making use of timing information, it is possible to significantly reduce the systematic uncertainties and to independently measure the fluxes of neutrinos that originate as $\nu_{\mu}$ or as either $\nu_e$ or $\bar{\nu}_{\mu}$. We find that a 100 kg CsI detector would be sensitive to the large fraction of the sterile neutrino parameter space that could potentially account for the LSND and MiniBooNE anomalies.

\end{abstract}

\pacs{Valid PACS appear here}
\maketitle


\section{\label{sec:1}Introduction}

A large number of neutrino oscillation experiments carried out over the past two decades have lead to the establishment of a standard framework, in which the neutrino sector is comprised of three massive neutrinos with relatively large mixing angles and sub-electronvolt mass splittings. Some of the data that has been collected, however, cannot be explained within the context of this standard framework. Particularly puzzling have been the results of the Liquid Scintillator Neutrino Detector (LSND)~\cite{Aguilar:2001ty} and MiniBooNE~\cite{Aguilar-Arevalo:2018gpe} experiments. In the former, an excess of $\bar\nu_e$ events was observed from a source of pions decaying at rest, while in the latter an excess of electron-like events was observed in both $\nu_\mu$ and $\bar\nu_\mu$ beams. Although these experiments bear very distinct neutrino energy profiles, their results are each consistent with $\nu_\mu\to\nu_e$  oscillations occurring over a short baseline, requiring a relatively large mass splitting~\cite{Dentler:2018sju}. Moreover, reexaminations~\cite{Bahcall:1997eg, Acero:2007su, Giunti:2010zu, Kopp:2013vaa} of the Gallium calibration experiments~\cite{Hampel:1998xg, Abdurashitov:2005ax} and reactor antineutrino fluxes~\cite{Mention:2011rk, Huber:2011wv} have identified a deficit of $\nu_e$ events relative to theoretical expectations.

Together, these anomalies appear to point to the existence of a sterile neutrino with a mass near the electronvolt scale. Complicating this interpretation, however, are the results of several experiments which measure $\nu_\mu\to\nu_\mu$ transitions. In particular, the IceCube~\cite{TheIceCube:2016oqi} and MINOS/MINOS+~\cite{Adamson:2017uda} experiments should be sensitive to the presence of such a sterile neutrino, but have observed no deviation from the standard (three-neutrino) framework. Detailed statistical analyses of the combined short baseline neutrino data have revealed a large degree of tension, at the $4.7\sigma$ level, between the appearance ($\nu_\mu\to\nu_e$) and disappearance ($\nu_\mu\to\nu_\mu$ and $\nu_e\to\nu_e$) data sets, at least within the context of oscillations with one or more sterile neutrinos~\cite{Collin:2016rao, Gariazzo:2017fdh, Dentler:2018sju}.

Motivated by this tension, a number of alternative explanations for the short baseline anomalies have recently been put forward. For example, models in which neutrinos upscatter to heavier fermions can provide a good fit to MiniBooNE's energy spectrum and angular distribution~\cite{Bertuzzo:2018itn, Bertuzzo:2018ftf, Ballett:2018ynz} (see also \cite{Arguelles:2018mtc}), but cannot account for the LSND excess. Other explanations involving ``dark'' particles produced in the beam have also been considered, but are strongly constrained by the MiniBooNE beam dump run data~\cite{Aguilar-Arevalo:2018wea, Jordan:2018qiy}. At this time, the correct interpretation of this data remains unclear, and further information will be required in order to clarify this confusing situation.

In this letter, we propose to measure or constrain the occurrence of short-baseline active-to-sterile neutrino oscillations using neutrinos from a source of pions decaying at rest, which are then measured via neutral current coherent elastic neutrino-nucleus scattering. This interaction, which has recently been observed for the first time by the COHERENT Collaboration~\cite{Akimov:2017ade}, using a CsI target~\cite{COLLAR201556}, is universal across all three active neutrino flavors, and thus can provide a measurement of the total (all flavor) active neutrino flux~\cite{Anderson:2012pn,Kosmas:2017zbh} (see also Refs.~\cite{Dutta:2015nlo,Canas:2017umu,Formaggio:2011jt}). Furthermore, by taking data on two different baselines, we can substantially reduce the systematic uncertainties associated with the detector efficiency, as well as with the overall flux and cross section. The use of timing information provides information pertaining to neutrino flavor, allowing us to disentangle the effects of sterile neutrino mixing with $\nu_\mu$ and $\nu_e$, and thus providing a powerful probe of the LSND and MiniBooNE anomalies.



\section{\label{sec:2} Coherent Neutrino Scattering}

For our experimental setup, we have in mind a source of neutrinos that is functionally similar to the Spallation Neutron Source at the Oak Ridge National Laboratory~\cite{Avignone:2003ep,Akimov:2015fsi}. In particular, we consider a pulsed beam of 1 GeV protons with a luminosity of $L_0=4\times 10^{23}$ yr$^{-1}$, of which approximately 8\% (6\%) produce a $\pi^+$ ($\pi^-$) upon striking the spallation target.

The high-$Z$ nature of the liquid mercury target depletes virtually all of the negatively charged pions through nuclear capture (only $\sim 2.3\times10^{-5}$ decay prior to nuclear capture), while the positively charged pions are rapidly slowed but not captured. These particles then decay at rest, via $\pi^+ \rightarrow \mu^+ \nu_{\mu} \rightarrow e^+ \nu_e  \bar{\nu}_{\mu} \nu_{\mu}$, yielding the following isotropic spectrum per $\pi^+$:
\begin{align}
 \frac{dN_{\nu_{\mu}}}{dE_{\nu_{\mu}}}&=\delta \left( E_{\nu_{\mu}}-\left(\frac{m^2_\pi - m^2_\mu}{2 m_\pi} \right)\right), \\
\frac{dN_{\bar\nu_{\mu}}}{dE_{\bar\nu_{\mu}}}&=\frac{6 E_{\bar\nu_{\mu}}^2}{(m_\mu / 2)^4} \left( \frac{m_\mu}{2} - \frac{2}{3}E_{\bar\nu_{\mu}} \right), \; E_{\bar\nu_{\mu}} \leq m_\mu/2, \nonumber \\
\frac{dN_{\nu_e}}{dE_{\nu_e}}&=\frac{12 E_{\nu_e}^2}{(m_\mu / 2)^4} \left( \frac{m_\mu}{2} - E_{\nu_e} \right), \; E_{\nu_e} \leq m_\mu/2, \nonumber
\end{align}
where $m_\mu$ and $m_\pi$ are the masses of the muon and pion, respectively. 

After travelling to the location of the detector, these neutrinos can be observed through coherent elastic neutrino-nucleus scattering. The cross section for this process is given as follows:
\begin{multline}
\frac{d\sigma}{dE_T} = \frac{G^2_F M}{4\pi} [N - Z\left(1-4 \sin^2 \theta_{\rm W} \right)]^2\times \\
\left( 1- \frac{E_T}{E_\nu} - \frac{M E_T}{2 E^2_\nu} + \frac{E^2_T}{2E^2_\nu}\right)F^2(E_T),
\label{crosssection}
\end{multline}
where $E_T$ is the recoil energy, $E_\nu$ is the neutrino energy, $\theta_{\rm W}$ is the weak mixing angle, $M$ is the target nuclear mass, $N$ $(Z)$ is the number of neutrons (protons) in the nucleus, and $F$ is the Helm nuclear form factor given by the following~\cite{Lewin:1995rx}:
\begin{align}
   F(Q)&=\frac{3j_1(R_0 Q)}{R_0 Q} \exp{\left( -\frac{Q s}{2}\right)}\\
   j_1(x)&=\frac{\sin{x}}{x^2}-\frac{\cos{x}}{x}, \nonumber
\end{align}
where $j_1$ is the first-order spherical-Bessel function, $Q^2=-p_\mu p^\mu=2ME_T$ is the momentum transfer, and $R^2_0=\sqrt{R^2-5s^2}$ is determined by the surface thickness, $s=0.5\left( \text{fm} \right)$, and the effective nuclear radius, $R=1.2A^{\frac{1}{3}}\left( \text{fm} \right)$~\cite{Kosmas:2017tsq}. In this study, we will focus on the case of CsI as our target for coherent scattering.


The total coherent scattering cross section is given by integrating Eq.~\ref{crosssection} up to a maximum recoil energy of $E_{Tmax}=\left( 2 E^2_\nu / (M+2E_\nu)\right)$. We also include a signal acceptance function, which we take to interpolate linearly between zero at $E_T=$2.5 keV and 100\% at $E_T=$5 keV. This represents only a small improvement over that described in Ref.~\cite{Akimov:2017ade}.


\begin{figure}[t]
    \centering
    \includegraphics[width=86mm]{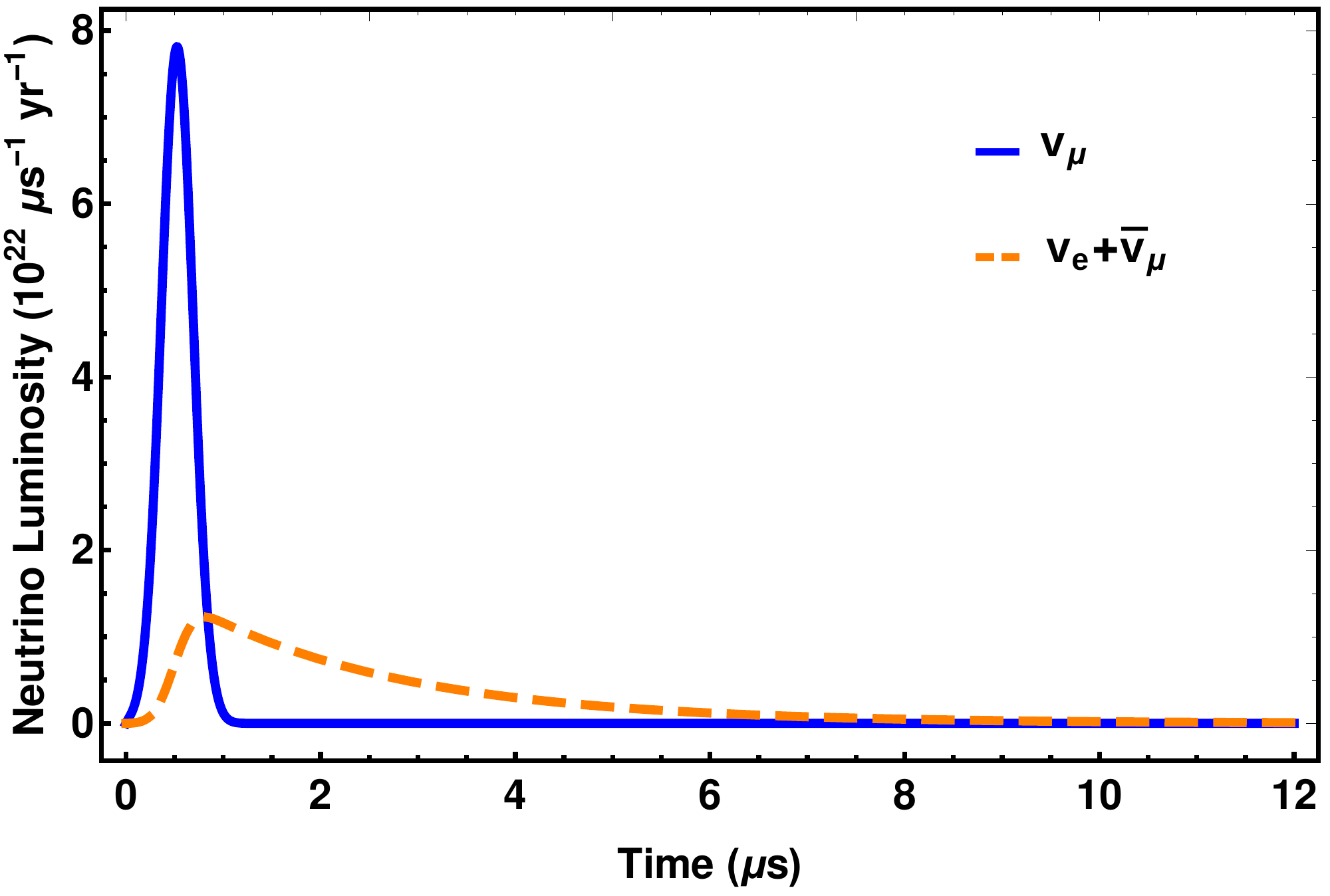}
    \caption{The neutrino luminosity from a pulsed source of charged pions decaying at rest. The muon lifetime ($\tau_{\mu} = 2.197 \, \mu$s) delays the $\nu_e$ and $\bar{\nu}_{\mu}$ emission to a degree that is long compared to the pion lifetime ($0.026 \, \mu$s) and pulse width ($0.16 \, \mu$s), but is very short compared to the time between pulses at the Spallation Neutron Source (0.017 s). This makes it possible to use timing information to separate the $\nu_{\mu}$ flux from the combined flux of $\nu_e$ and $\bar{\nu}_{\mu}$. Note that we have defined the time-axis such that $t=0.5\,\mu$s is at the center of the pulse.}
    \label{fig:3}
\end{figure}

By utilizing timing information, it is possible to distinguish the $\nu_{\mu}$ flux from the combined flux of $\nu_e$ and $\bar{\nu}_{\mu}$. In particular, the muon lifetime ($\tau_{\mu} = 2.197 \, \mu$s) delays the $\nu_e$ and $\bar{\nu}_{\mu}$ arrival times to a degree that is long compared to both the pion lifetime ($0.026 \, \mu$s) and the Gaussian pulse width ($0.16 \, \mu$s), but short compared to the time between pulses at the Spallation Neutron Source (0.017 s). This is illustrated in Fig.~\ref{fig:3}, where we show the time profile of the neutrino luminosity from such a pulsed spallation source.

\section{Projected Sensitivity to Sterile Neutrinos}

\begin{figure*}[t]
    \centering
    \includegraphics[width=88mm]{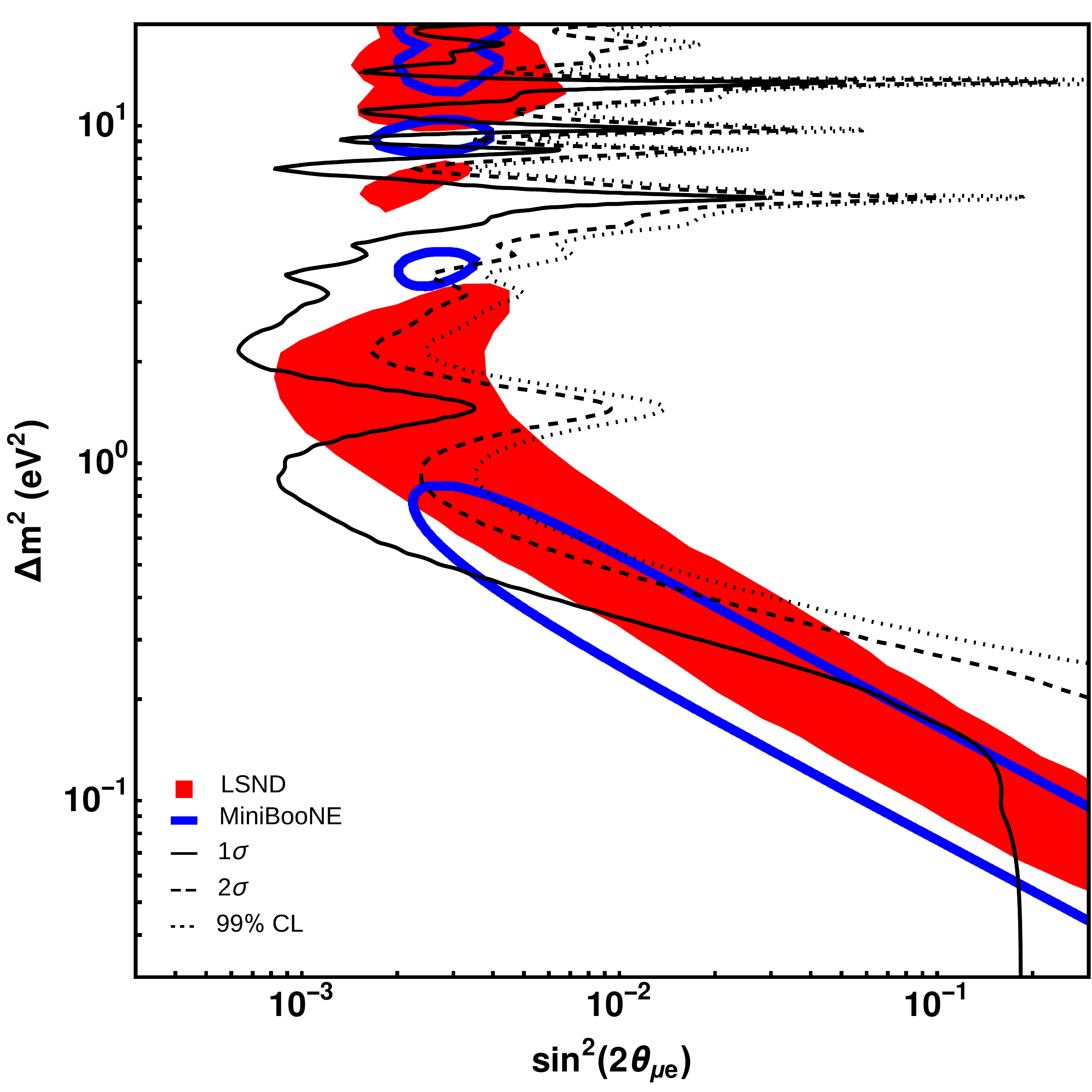}
    \includegraphics[width=88mm]{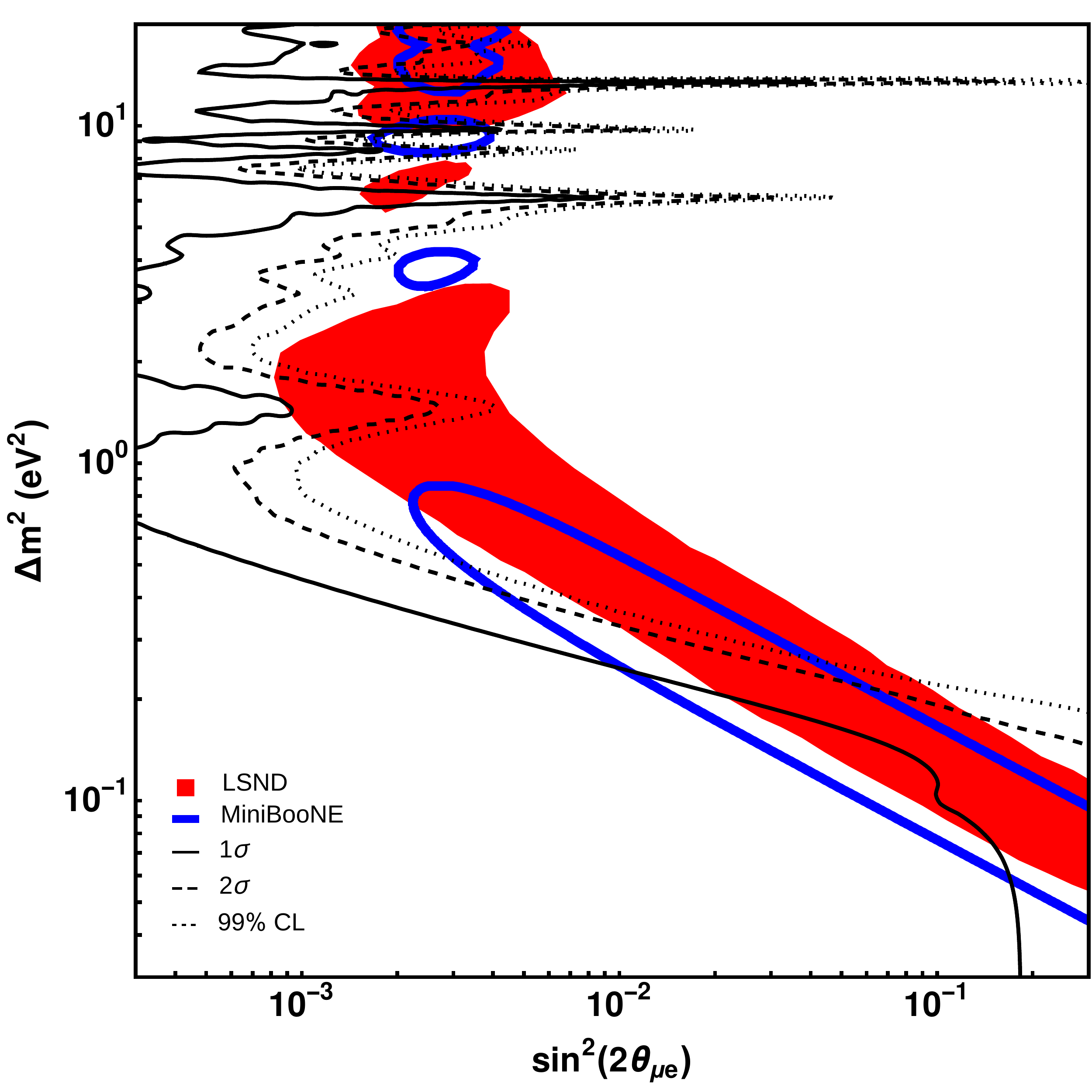}
    \caption{The projected constraints on the sterile neutrino parameter space for a 100 kg CsI detector and source that generates $4\times 10^{23}$ protons on target per year with an energy of 1 GeV, after collecting data for a total of 3 years (left) or 10 years (right). In each case, we have assumed that the detector was located at a distance of 20 meters from the source during the first half of the exposure, and at a distance of 40 meters during the second half. These constraints are compared to the regions that could potentially account for the LSND~\cite{Aguilar:2001ty} and MiniBooNE~\cite{Aguilar-Arevalo:2018gpe} anomalies (at the 99\% confidence level).}
    \label{fig:2}
\end{figure*}

The probability of a neutrino of flavor $\alpha$ oscillating to flavor $\beta$ is given by the following:
 \begin{equation}
     P_{\alpha \rightarrow \beta} = \sum_{j,k} U^*_{\alpha,j} U_{\beta,j} U_{\alpha,k} U^*_{\beta,k}\exp{\left(-i \frac{\Delta m^2_{jk} L}{2E_{\nu}}\right)},
     \label{prob}
 \end{equation}
 where $L$ is the distance travelled, $U$ is the neutrino mixing matrix, $\Delta m^2_{jk}=m^2_j - m^2_k$ is the mass squared difference between two neutrino mass eigenstates, and $E_{\nu}$ is the neutrino energy. Given that we are interested in the case in which the three Standard Model neutrinos are each much lighter than the sterile state ($m_{1,2,3} \ll m_4$), we can simplify the problem by setting $\Delta m^2_{41} \approx \Delta m^2_{42} \approx \Delta m^2_{43}$. Furthermore, Eq.~\ref{prob} simplifies considerably on short baselines, for which oscillations between Standard Model (non-sterile) neutrinos are negligible. In this regime, the probability of a neutrino of flavor $\alpha$ arriving at the detector as an active neutrino is given as follows:
\begin{eqnarray}
P_{\alpha \rightarrow e, \mu, \tau} &=&  1 - 4 |U_{\alpha 4}|^2 \bigg(1-\sum_{\beta=e,\mu,\tau} |U_{\beta 4}|^2\bigg) \sin^2\bigg(\frac{\Delta m^2 L}{4E}\bigg), \nonumber \\
\end{eqnarray}
 where $\Delta m^2 \equiv \Delta m^2_{41} \approx  \Delta m^2_{42} \approx \Delta m^2_{43}$. After appropriate unit conversions, the argument of the sine function can be rewritten as $1.27 \times (\Delta m^2/{\rm eV}^2) (L/{\rm m}) ({\rm MeV}/E)$.

 To estimate the sensitivity of a coherent scattering experiment to a sterile neutrino, we calculate the number of events predicted to be observed in four time bins, corresponding to 0-0.5, 0.5-1, 1-2 and 2-10 $\mu$s, defined such that $t=0.5\,\mu$s is at the center of the pion pulse. By taking into account this timing information, we are able to measure independently the flux of neutrinos that originate as $\nu_{\mu}$, as well and those that originate as either $\nu_e$ or $\bar{\nu}_{\mu}$. We consider measurements made over two baselines (20 and 40 meters), using a target consisting of 100 kg of CsI, over a total observation time of either 3 or 10 years (half of the total time is assumed to be in each of the 20 and 40 meter configurations). In each configuration, we include in the event rate calculation a steady-state background of $1.45\times 10^{-10}$ counts/kg/$\mu$s, which can be precisely determined by measuring the off-pulse event rate~\cite{Akimov:2017ade}. Although this is a factor of 10 lower than the rate reported in Ref.~\cite{Akimov:2017ade}, this degree of improvement is achievable through the application of additional shielding and well-understood techniques to reduce the dominant internal radiocontaminations of CsI \footnote{Juan Collar, private communication.} (see also Refs.~\cite{Kim:2008zzp,Kim:2003ms,Kim:2005rr}). We also include in our analysis an overall systematic uncertainty of $\pm28\%$ on the overall signal rate, corresponding to uncertainties associated with the cross section, detector efficiency, and overall neutrino flux.

\begin{figure*}
    \centering
\includegraphics[width=88mm]{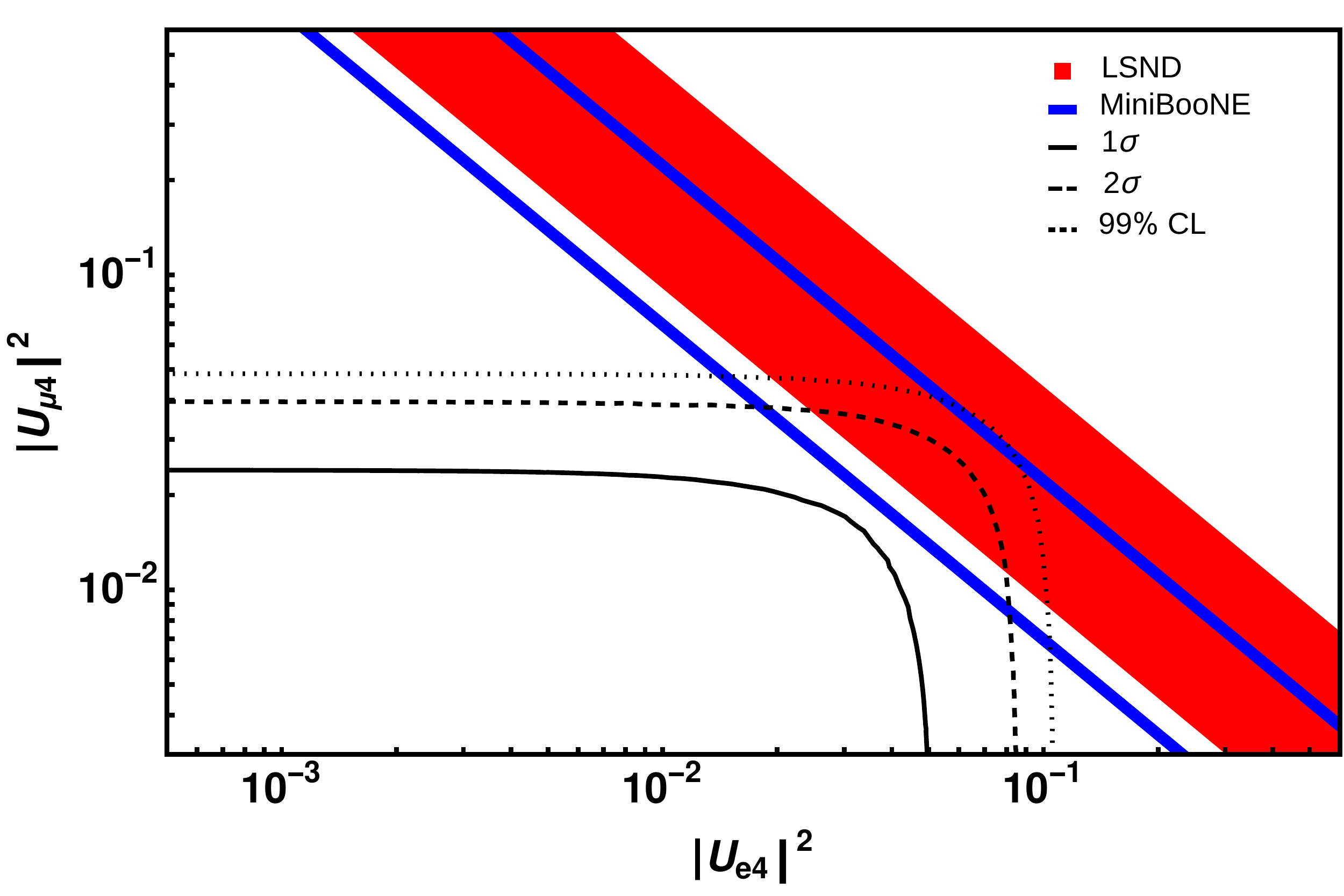}
\includegraphics[width=88mm]{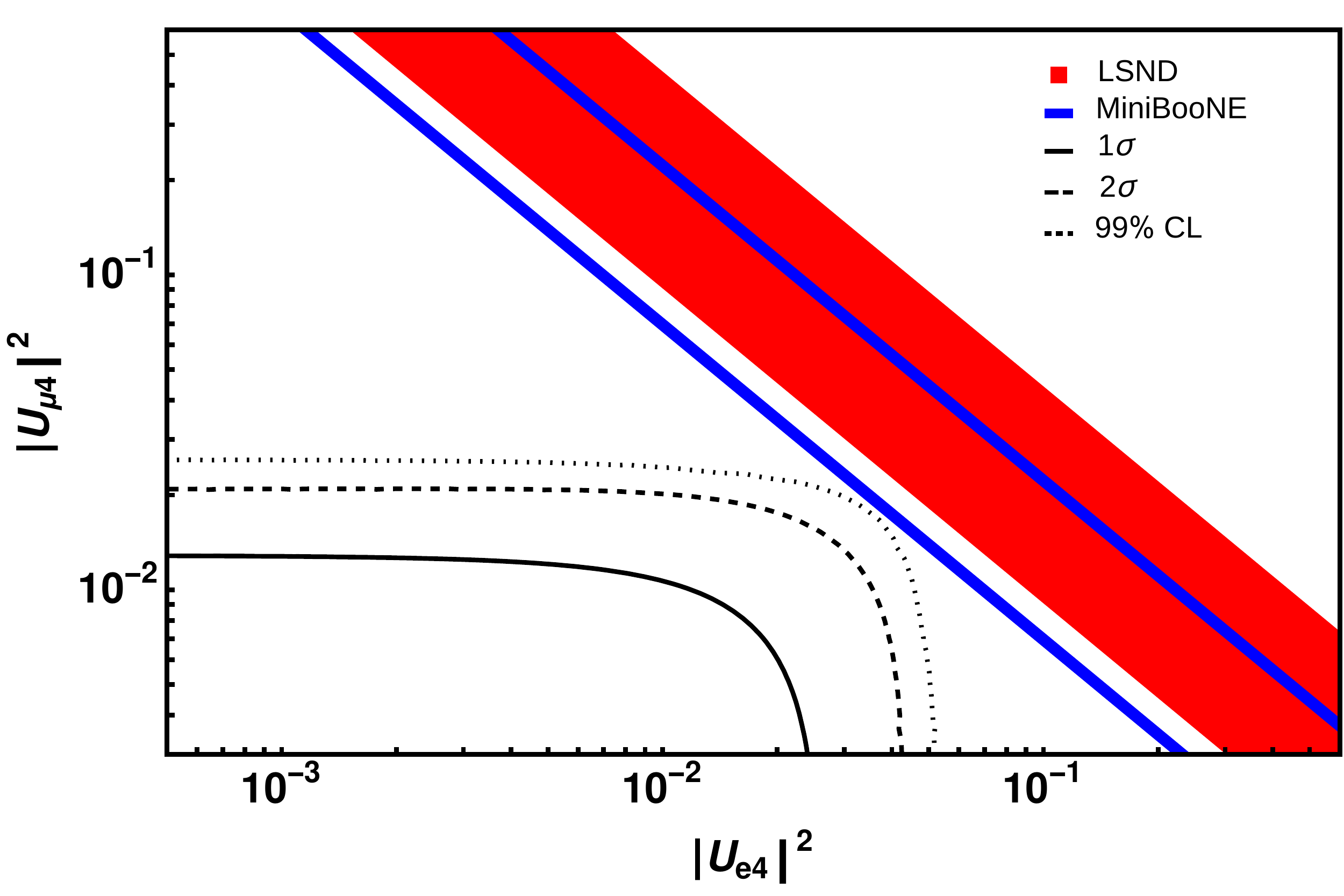}\\
\includegraphics[width=88mm]{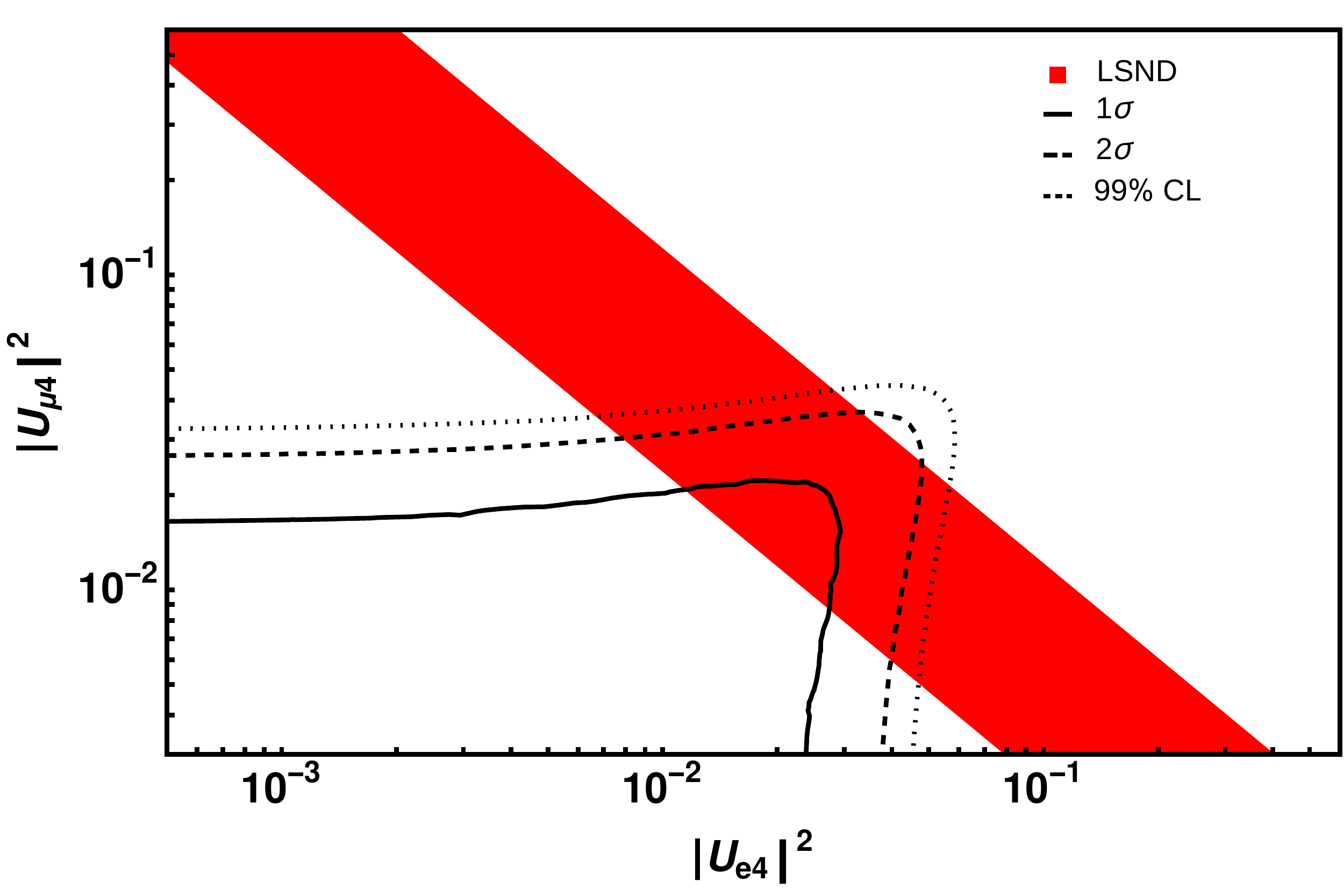}
\includegraphics[width=88mm]{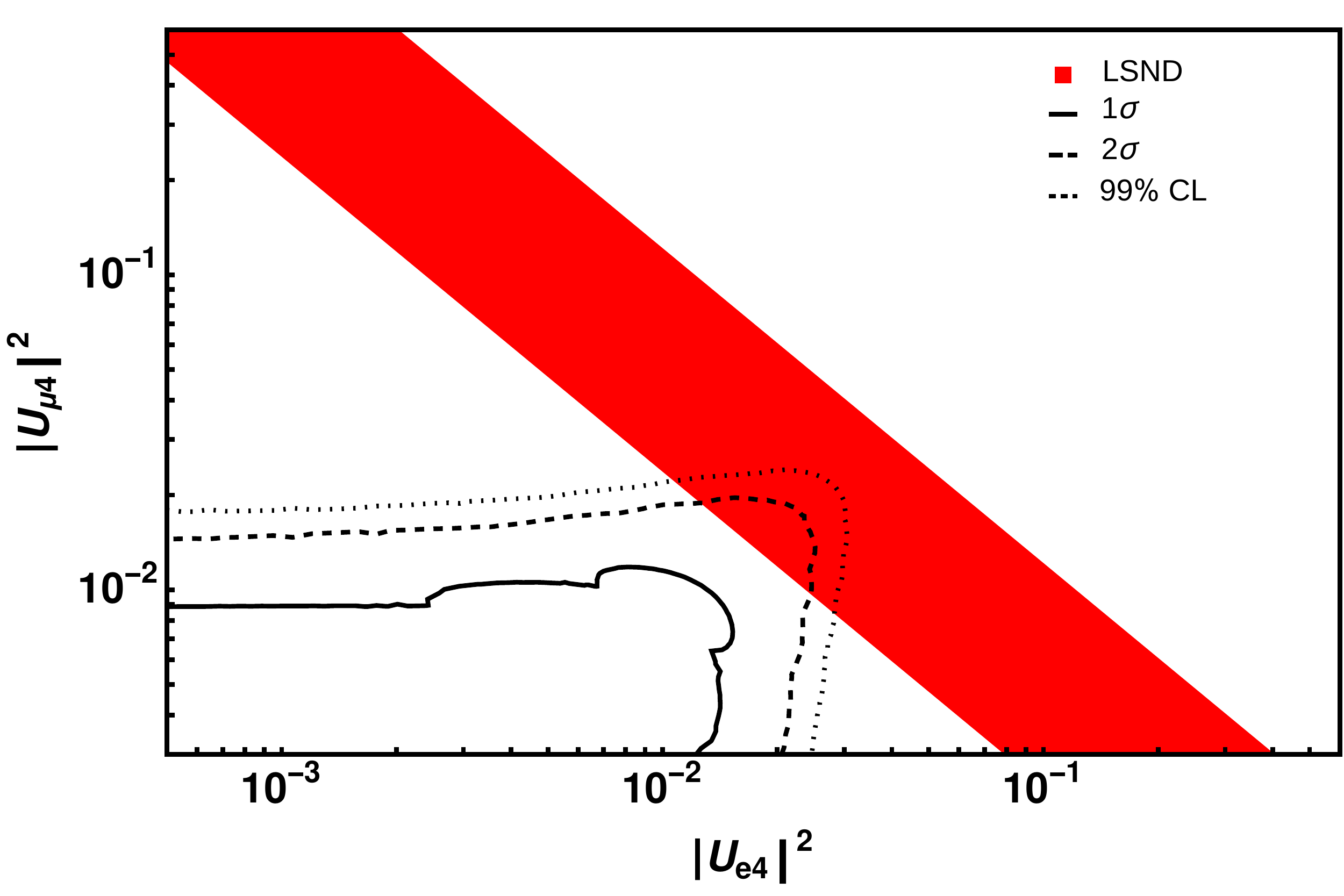}
    \caption{As in Fig.~\ref{fig:2}, but in terms of the $|U_{\mu4}|^2$ vs $|U_{e4}|^2$ parameter space, for the choice of either $\Delta m^2=0.55$ eV$^2$ (upper frames) or $\Delta m^2=1.3$ eV$^2$ (lower frames). Again, the left (right) frames are after collecting data for 3 (10) years. These constraints are compared to the regions that could potentially account for the LSND~\cite{Aguilar:2001ty} and MiniBooNE~\cite{Aguilar-Arevalo:2018gpe} anomalies (at the 99\% confidence level).}
    \label{fig:4}
\end{figure*}

 The main results of our analysis are shown in Figs.~\ref{fig:2} and~\ref{fig:4}. In the first of these figures, we show the projected constraints on the sterile neutrino parameter space from an experiment utilizing a 100 kg CsI detector and source producing a luminosity of $4\times 10^{23}$ protons on target per year with an energy of 1 GeV. We present this result in terms of the effective mixing parameter $\sin^2 (2\theta_{\mu e}) \equiv 4 |U_{e4}|^2 |U_{\mu 4}|^2$. For simplicity, we have limited our discussion to the case of $|U_{\tau 4}| =0$. In Fig.~\ref{fig:4}, we show these constraints in the $|U_{\mu4}|^2$ vs $|U_{e4}|^2$ plane, for two choices of $\Delta m^2$. When these constraints are compared to the regions favored by LSND and MiniBooNE, we conclude that the search proposed here would be sensitive to the vast majority of the sterile neutrino parameter space that could potentially account for these anomalies.

In these projections, we have considered measurements taken over baselines of 20 and 40 meters. If the separation between these distances were increased (decreased), the exclusion contours would shift downward (upward) in $\Delta m^2$, as a consequence of the dependence on $L$ and $\Delta m^2$ in Eq.~5.

\section{Summary and Conclusions} 
 
In this letter, we have proposed using a 100 kg CsI coherent neutrino detector located near a pulsed source of neutrinos functionally similar to the Spallation Neutron Source at the Oak Ridge National Laboratory to test sterile neutrino scenarios motivated by the LSND and MiniBooNE anomalies. By making use of timing information, one independently measure the fluxes of neutrinos that originate as $\nu_{\mu}$ or as either $\nu_e$ or $\bar{\nu}_{\mu}$. Furthermore, by comparing the coherent scattering rates observed by a given detector while positioned at multiple distances from the source, it is possible to significantly reduce systematic uncertainties associated with the flux normalization, coherent scattering cross section, and detector efficiencies. We find that such an experiment would be sensitive to much of the relevant parameter space and would help to clarify the nature of the mysterious results reported by the LSND and MiniBooNE Collaborations.

\bigskip

We would like to thank Juan Collar for helpful discussions. CB is supported by the US National Science Foundation Graduate Research Fellowship under grants number DGE-1144082 and DGE-1746045. This manuscript has been authored by Fermi Research Alliance, LLC under Contract No. DE-AC02-07CH11359 with the U.S. Department of Energy, Office of Science, Office of High Energy Physics.

\bibliography{CEvNS}

\end{document}